\documentclass[conference]{IEEEtran}
\IEEEoverridecommandlockouts
\usepackage{cite}
\usepackage{amsmath,amssymb,amsfonts}
\usepackage{url}
\usepackage{algorithmic}
\usepackage{graphicx}
\usepackage{textcomp}
\usepackage{enumitem}
\usepackage{booktabs}
\usepackage{siunitx}
\usepackage{multirow}
\usepackage{subfig}
\usepackage{xcolor}
\def\BibTeX{{\rm B\kern-.05em{\sc i\kern-.025em b}\kern-.08em
    T\kern-.1667em\lower.7ex\hbox{E}\kern-.125emX}}

\usepackage{siunitx}
\usepackage{enumitem}

\usepackage[acronym,shortcuts]{glossaries}
\newacronym{BF}{BF}{boundary F1}
\newacronym{ML}{ML}{machine learning}
\newacronym{AI}{AI}{artificial intelligence}
\newacronym{IoT}{IoT}{internet of things}
\newacronym{PID}{PID}{proportional-integral-derivative}
\newacronym{RTT}{RTT}{round-trip time}
\newacronym{SSIM}{SSIM}{structural similarity index measure}
\newacronym{VPU}{VPU}{vision processing unit}

\begin{document}

\title{Network-Adaptive Cloud Processing for Visual Neuroprostheses}

\author{\IEEEauthorblockN{Jiayi Liu}
\IEEEauthorblockA{\textit{Computer Science} \\
\textit{UC Santa Barbara}\\
Santa Barbara, CA, USA \\
\url{jiayi979@ucsb.edu}}
\and
\IEEEauthorblockN{Yilin Wang}
\IEEEauthorblockA{\textit{Computer Science} \\
\textit{UC Santa Barbara}\\
Santa Barbara, CA, USA \\
\url{yilin_wang@ucsb.edu}}
\and
\IEEEauthorblockN{Michael Beyeler}
\IEEEauthorblockA{\textit{Computer Science} \\
\textit{Psychological \& Brain Sciences} \\
\textit{UC Santa Barbara}\\
Santa Barbara, CA, USA \\
\url{mbeyeler@ucsb.edu}}
}

\maketitle

\begin{abstract}
Cloud-based machine learning is increasingly explored as a preprocessing strategy for next-generation visual neuroprostheses, where advanced scene understanding may exceed the computational and energy constraints of battery-powered visual processing units. Offloading computation to remote servers enables the use of state-of-the-art vision models, but also introduces sensitivity to network latency, jitter, and packet loss, which can disrupt the temporal consistency of the delivered neural stimulus. In this work, we examine the feasibility of cloud-assisted visual preprocessing for artificial vision by framing remote inference as a perceptually constrained systems problem. We present a network-adaptive cloud-assisted pipeline in which real-time round-trip-time feedback is used to dynamically modulate image resolution, compression, and transmission rate, explicitly prioritizing temporal continuity under adverse network conditions. PIDNet is used as a fixed real-time semantic segmentation backbone, allowing us to isolate how network-adaptive input encoding affects communication delay, inference time, and perceptual fidelity. Results show that adaptive visual encoding substantially reduces end-to-end latency during network congestion, with only modest degradation of global scene structure, while boundary precision degrades more sharply. Together, these findings delineate operating regimes in which cloud-assisted preprocessing may remain viable for future visual neuroprostheses and underscore the importance of network-aware adaptation for maintaining perceptual stability.
\end{abstract}


\begin{figure*}[!t]
    \centering
    \includegraphics[width=\linewidth]{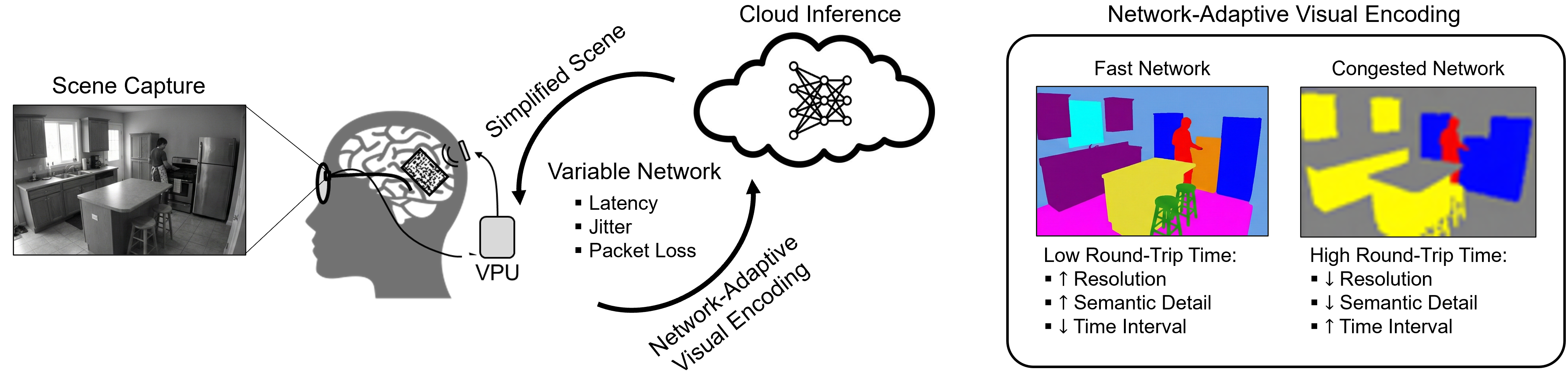}
    \caption{Network-adaptive cloud processing for visual neuroprostheses.
        Egocentric video captured by a resource-constrained \acf{VPU} is adaptively encoded prior to transmission based on real-time network feedback. \Acf{RTT} measurements drive a closed-loop controller that modulates image resolution, compression, and transmission interval to maintain temporal continuity of the delivered visual stimulus under network impairments. Remote semantic segmentation is performed in the cloud, and the resulting simplified scene is returned to the client. In the present system, reduced semantic detail under network congestion arises indirectly from input degradation; future implementations could achieve similar effects by explicitly requesting coarse vs. fine semantic outputs from the cloud inference service. Scene images were generated using Nano Banana Pro for illustrative purposes only.}
\label{fig:overview}
\end{figure*}

\section{INTRODUCTION}

Bionic vision is increasingly exploring the use of \ac{ML} and \ac{AI} as preprocessing tools for next-generation retinal and cortical prostheses~\cite{beyeler_towards_2022,de_ruyter_van_steveninck_end--end_2022,granley_hybrid_2022,granley_human---loop_2023}.
Rather than operating directly on raw camera input, many proposed approaches seek to transform egocentric video streams into simplified, task-relevant representations that may be more interpretable to the visual system. 
These strategies include semantic and structural edge extraction~\cite{sanchez_garcia_semantic_2020,han_deep_2021,de_ruyter_van_steveninck_real-world_2022}, object-level abstraction~\cite{kasowski_static_2025,nejad_point-spv_2025}, and depth-aware scene simplification~\cite{barnes_enhancing_2015,rasla_relative_2022}, with the goal of supporting mobility and navigation. 
While such \ac{AI}-based preprocessing is not yet deployed in current clinical devices, it represents a rapidly growing research direction for mitigating the perceptual limitations imposed by existing prosthetic hardware.

A major obstacle to these \ac{AI}-driven approaches is the severe hardware constraint of current clinical systems.
Contemporary prosthetic devices rely on battery-powered \acp{VPU} that prioritize energy efficiency and thermal safety over computational throughput. 
As a result, many modern \ac{ML} models, particularly those required for complex indoor scene understanding, exceed the feasible limits of onboard processing and cannot be deployed directly on the device.

To address this computational gap, two complementary research trajectories have emerged largely outside the bionic vision literature. 
One approach focuses on highly efficient, lightweight algorithms, often referred to as \emph{TinyML}, that are explicitly designed for low-power \ac{IoT}-class devices~\cite{warden_tinyml_2020}. 
Although promising, such models typically achieve efficiency by reducing representational depth or semantic richness, which can limit their effectiveness for visually demanding mobility tasks~\cite{banbury_benchmarking_2021,howard_mobilenets_2017}.
A second approach leverages cloud-based inference, offloading computationally intensive processing to remote servers capable of running state-of-the-art vision models~\cite{kang_neurosurgeon_2017,nan_large-scale_2023}. 
While this paradigm is widely used in other vision and sensing domains, it has received limited attention in bionic vision research, particularly with respect to its sensitivity to network latency, jitter, and packet loss~\cite{laskaridis_spinn_2020,zobaed_edge-multiai_2022}.

For visual neuroprostheses, the vulnerability of cloud-centric preprocessing pipelines to network-induced delays presents a fundamental challenge. 
Unlike conventional computer vision applications, delays in prosthetic vision are not merely an inconvenience but would directly affect the temporal consistency of the delivered neural stimulus~\cite{held_telepresence_1992,miall_adaptation_2006}.
Prior work in visual perception and sensorimotor integration indicates that delays on the order of tens to hundreds of milliseconds can degrade perceptual stability, impair visuomotor coordination, and increase disorientation during mobility tasks~\cite{miall_adaptation_2006,stetson_motor-sensory_2006,honda_adaptation_2012,beech_visuomotor_2025}. 
In the context of artificial vision, such delays may manifest as interruptions in perceptual continuity, undermining both usability and safety.

Here, we investigate cloud-based semantic segmentation as a candidate preprocessing strategy for future visual neuroprostheses and address its central limitation by introducing a network-adaptive control policy (Figure~\ref{fig:overview}). 
Our approach dynamically modulates visual encoding parameters based on real-time network feedback, explicitly prioritizing temporal continuity of the percept over spatial fidelity as network conditions degrade. 
By framing cloud inference as a perceptually constrained control problem rather than a static vision pipeline, we seek to identify operating regimes in which cloud-based preprocessing can remain viable for artificial vision despite stochastic network impairments.

Our contributions are threefold:
\begin{enumerate}[topsep=0pt,itemsep=-1ex,partopsep=0pt,parsep=1ex,leftmargin=14pt,label=\roman*.]
    \item We present a cloud-assisted visual preprocessing pipeline tailored to the constraints of visual neuroprostheses and systematically characterize its sensitivity to network impairments across a range of realistic connectivity regimes. 
    \item We introduce a closed-loop, network-adaptive encoding strategy that explicitly trades spatial fidelity for temporal stability, enabling the system to maintain a minimum perceptual update rate under adverse network conditions. 
    \item We quantify the resulting latency-fidelity trade-off using perceptually motivated measures, demonstrating that substantial reductions in end-to-end delay can be achieved with only modest degradation of global scene structure, while boundary precision degrades more sharply under severe congestion.
\end{enumerate}
Together, this work establishes a principled framework for evaluating when and how cloud-assisted visual preprocessing can be integrated into future neuroprosthetic systems without violating perceptual timing constraints critical for safe and effective use.

\section{METHODS}

\subsection{System Overview and Architecture}

Figure~\ref{fig:overview} provides an overview of the proposed network-adaptive cloud-assisted visual preprocessing system. 
The architecture is designed to approximate a future visual neuroprosthesis workflow in which a resource-constrained \acf{VPU} captures egocentric video and offloads computationally intensive preprocessing to a remote server. 
The central challenge addressed by this system is maintaining perceptual temporal continuity in the presence of stochastic network impairments.

The system comprises three principal components: 
\begin{enumerate}[topsep=0pt,itemsep=-1ex,partopsep=0pt,parsep=1ex,leftmargin=14pt,label=\roman*.]
    \item a \ac{VPU}-side client responsible for video capture and adaptive visual encoding, 
    \item a bidirectional network channel subject to controlled latency, bandwidth, and packet loss, and 
    \item a cloud-based inference server that performs semantic preprocessing and returns a simplified scene representation to the client.
\end{enumerate}
A closed-loop control policy operating on the client dynamically adjusts visual encoding parameters based on real-time network feedback, explicitly trading spatial fidelity for reduced end-to-end delay under adverse network conditions.

\subsection{Network-Adaptive Visual Encoding Policy}

The core contribution of this work is a closed-loop, network-adaptive encoding policy that stabilizes the temporal delivery of cloud-based visual preprocessing for artificial vision. 
Rather than treating remote inference as a static computer vision pipeline, the proposed system explicitly accounts for network variability by regulating the visual encoding process on the \ac{VPU} in response to observed communication delays.

The design objective of the controller is to maintain perceptual temporal continuity of the delivered stimulus, even when network conditions degrade, by preventing excessive end-to-end latency accumulation. 
To achieve this, the controller dynamically modulates visual encoding parameters, prioritizing timely stimulus updates over spatial detail during periods of congestion.

\subsubsection{Network Feedback Signal}

Network responsiveness is quantified on the client using the \ac{RTT} between the \ac{VPU} and the cloud server. 
A dedicated monitoring thread periodically probes the communication channel and records the most recent $K$ \ac{RTT} measurements in a bounded buffer. 
To reduce sensitivity to transient jitter and outliers, the controller operates on a moving average estimate,
\begin{equation}
\overline{\mathrm{RTT}} = \frac{1}{K}\sum_{i=1}^{K}\mathrm{RTT}_i,
\end{equation}
where $K=5$ in the present implementation. 
This smoothed estimate serves as the sole feedback signal driving adaptive reconfiguration.

\subsubsection{Tiered Reconfiguration Policy}

Based on the estimated $\overline{\mathrm{RTT}}$, the controller selects one of five discrete operating regimes. 
Each regime specifies a visual encoding parameter vector
\[
\mathcal{P} = \{Q, R, I\},
\]
where $Q$ denotes the JPEG compression quality, $R$ the maximum image resolution (with aspect ratio preserved), and $I$ the inter-frame transmission interval.

Discrete operating tiers were chosen to ensure predictable and stable behavior under fluctuating network conditions while minimizing computational overhead on the \ac{VPU}. 
As network delay increases, the controller progressively reduces spatial resolution and compression quality and increases the inter-frame interval, thereby limiting queue buildup and preventing excessive end-to-end latency.

The thresholds and corresponding parameter settings defining each operating regime are summarized in Table~\ref{tab:adaptation_tiers}.

\begin{table}[t!]
    \centering
    \caption{Network-adaptive encoding tiers used by the closed-loop controller.}
    \renewcommand{\arraystretch}{1.2}
    \begin{tabular}{lccc}
        \toprule
        \textbf{RTT threshold} 
        & \textbf{JPEG quality} 
        & \textbf{Max. resolution} 
        & \textbf{Send interval} \\
        \midrule
        $\leq \SI{30}{\milli\second}$  & 90\% & 1920 px & \SI{80}{\milli\second}  \\
        $\leq \SI{50}{\milli\second}$  & 80\% & 1280 px & \SI{100}{\milli\second} \\
        $\leq \SI{100}{\milli\second}$ & 65\% & 960 px  & \SI{150}{\milli\second} \\
        $\leq \SI{150}{\milli\second}$ & 50\% & 720 px  & \SI{250}{\milli\second} \\
        $> \SI{150}{\milli\second}$    & 40\% & 480 px  & \SI{500}{\milli\second} \\
        \bottomrule
    \end{tabular}
    \label{tab:adaptation_tiers}
\end{table}

\subsection{Remote Semantic Preprocessing}

Semantic preprocessing is performed on the cloud server using PIDNet~\cite{xu_pidnet_2023}, a real-time semantic segmentation architecture inspired by proportional-integral-derivative control principles. 
PIDNet combines three complementary processing branches: 
(i) a proportional branch that preserves high-resolution spatial structure, 
(ii) an integral branch that aggregates global contextual information, and 
(iii) a derivative branch that emphasizes high-frequency boundary features.

PIDNet was selected because its design matches the constraints of cloud-assisted artificial vision: it is intended for real-time semantic segmentation, explicitly preserves boundary information, and provides a favorable trade-off between segmentation quality and inference speed. 
This is important for visual neuroprosthetic preprocessing, where the output need not be photorealistic but should preserve task-relevant scene layout and salient object boundaries. 
Compared with heavier encoder--decoder or transformer-based segmentation architectures, PIDNet is better suited to latency-sensitive deployment because it avoids the computational cost of very large segmentation backbones. 
Compared with highly compact mobile segmentation networks, PIDNet retains an explicit boundary-processing pathway, which is relevant because object and obstacle boundaries are often among the most useful visual features for prosthetic vision. 
Thus, PIDNet was used as a representative real-time segmentation backbone rather than as the object of an architecture benchmark.
Alternative choices include lightweight mobile segmentation networks, such as MobileNet-based~\cite{howard_mobilenets_2017} or BiSeNet-style models~\cite{yu_bisenet_2018}, and larger encoder--decoder or transformer-based models, such as SegFormer-style~\cite{xie_segformer_2021} or Mask2Former-style architectures~\cite{cheng_masked-attention_2022}.

The server processes each received frame independently and returns a simplified scene representation to the client as part of the end-to-end loop. 
In all experiments, the PIDNet architecture and learned parameters were held fixed. 
Performance differences across network conditions therefore arise from the interaction between network state and adaptive input encoding, rather than from changes to the segmentation model itself. 
Specifically, reducing image resolution and JPEG quality decreases the transmitted payload size and the number of input pixels processed by the server, which can reduce both communication delay and per-frame inference time while degrading spatial detail in the returned segmentation.

\subsection{Client--Server Communication}

Client and server communicate via gRPC~\footnote{\url{https://grpc.io}} over HTTP/2, enabling low-latency bidirectional request–response interactions using Protocol Buffers for binary serialization. 
Each video frame is transmitted as an encoded byte payload following application of the adaptive policy. 
Upon receipt, the server decodes the payload, performs semantic preprocessing, and returns the resulting representation to the client.

This request-response cycle constitutes a single iteration of the closed-loop system and is used as the basis for all latency and fidelity measurements reported in this study.

\subsection{Experimental Setup and Network Impairment Model}

System performance was evaluated under controlled network conditions designed to approximate a range of realistic wireless connectivity regimes. 
Network impairments were imposed on the server side using a network emulation framework that independently constrains uplink and downlink bandwidth while injecting additional latency and packet loss. 
This approach allows systematic evaluation of the proposed adaptive policy under reproducible yet heterogeneous network conditions.

We considered five network scenarios spanning severely constrained 4G-like connectivity to near-ideal 5G-like conditions (Table~\ref{tab:network_conditions}). 
These scenarios were selected to cover operating regimes relevant to mobile and wearable assistive technologies, including cases in which cloud-based preprocessing may become marginal or infeasible without adaptation.

For each network scenario, system performance was evaluated under two operating modes:
\textit{(i) a static baseline}, in which visual frames are transmitted at fixed resolution, compression quality, and frame rate; and
\textit{(ii) the proposed adaptive policy}, in which the encoding parameter vector $\mathcal{P}=\{Q,R,I\}$ is updated online based on the estimated $\overline{\mathrm{RTT}}$.

\begin{table}[t!]
    \centering
    \caption{Simulated network conditions used to evaluate cloud-assisted preprocessing.}
    \renewcommand{\arraystretch}{1.2}
    \begin{tabular}{lcccc}
        \toprule
        \textbf{Scenario} 
        & \textbf{Downlink} 
        & \textbf{Uplink} 
        & \textbf{RTT} 
        & \textbf{Loss} \\
        \midrule
        Extreme congested 4G & 10 Mbps  & 5 Mbps   & \SI{100}{\milli\second} & 5\%   \\
        Congested 4G         & 25 Mbps  & 10 Mbps  & \SI{100}{\milli\second} & 2\%   \\
        4G--5G hybrid        & 50 Mbps  & 25 Mbps  & \SI{50}{\milli\second}  & 0.5\% \\
        Good 5G              & 200 Mbps & 50 Mbps  & \SI{30}{\milli\second}  & 0.1\% \\
        Ultra-smooth 5G      & 800 Mbps & 200 Mbps & \SI{10}{\milli\second}  & 0\%   \\
        \bottomrule
    \end{tabular}
    \label{tab:network_conditions}
\end{table}

\begin{figure*}[!t]
    \centering
    \subfloat[Extreme Congested 4G]{
        \includegraphics[width=0.19\textwidth]{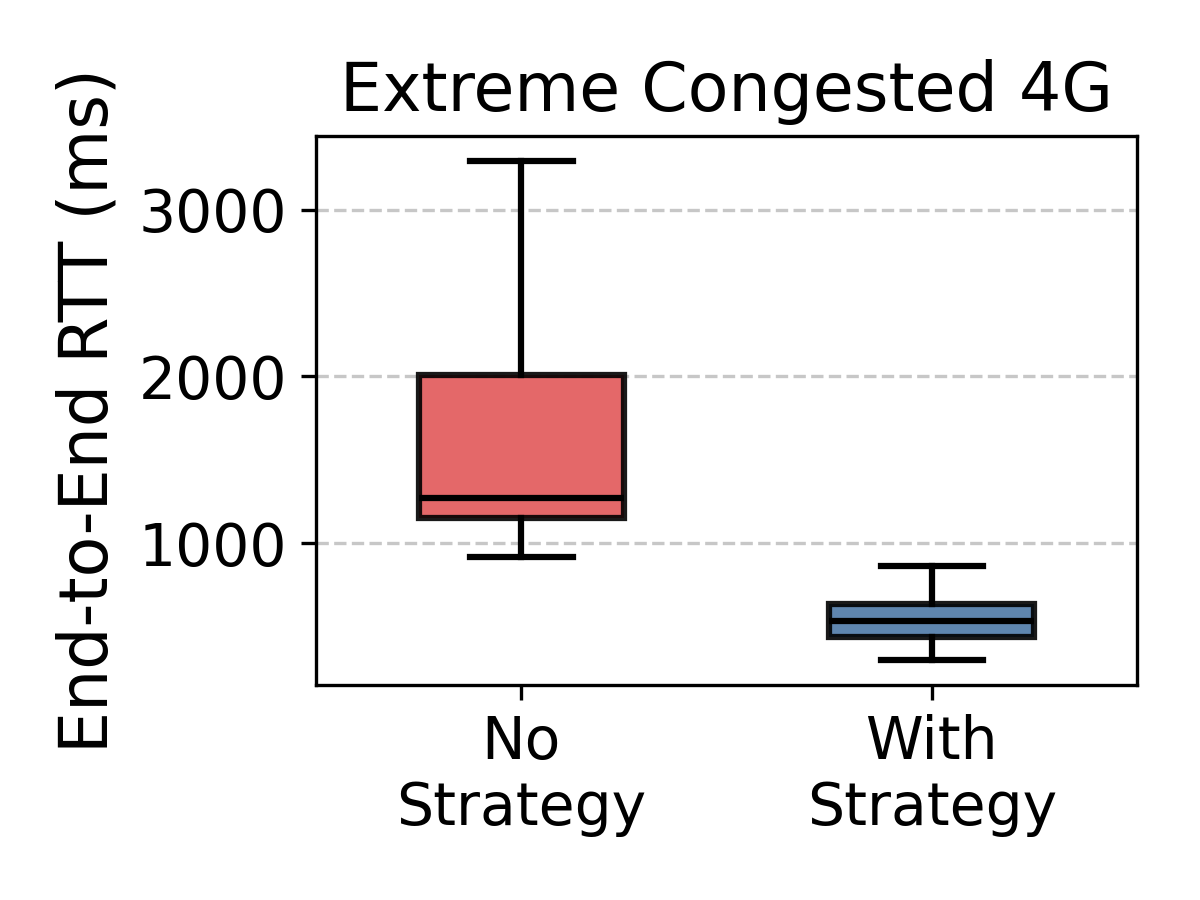}
    }
    \subfloat[Congested 4G]{
        \includegraphics[width=0.19\textwidth]{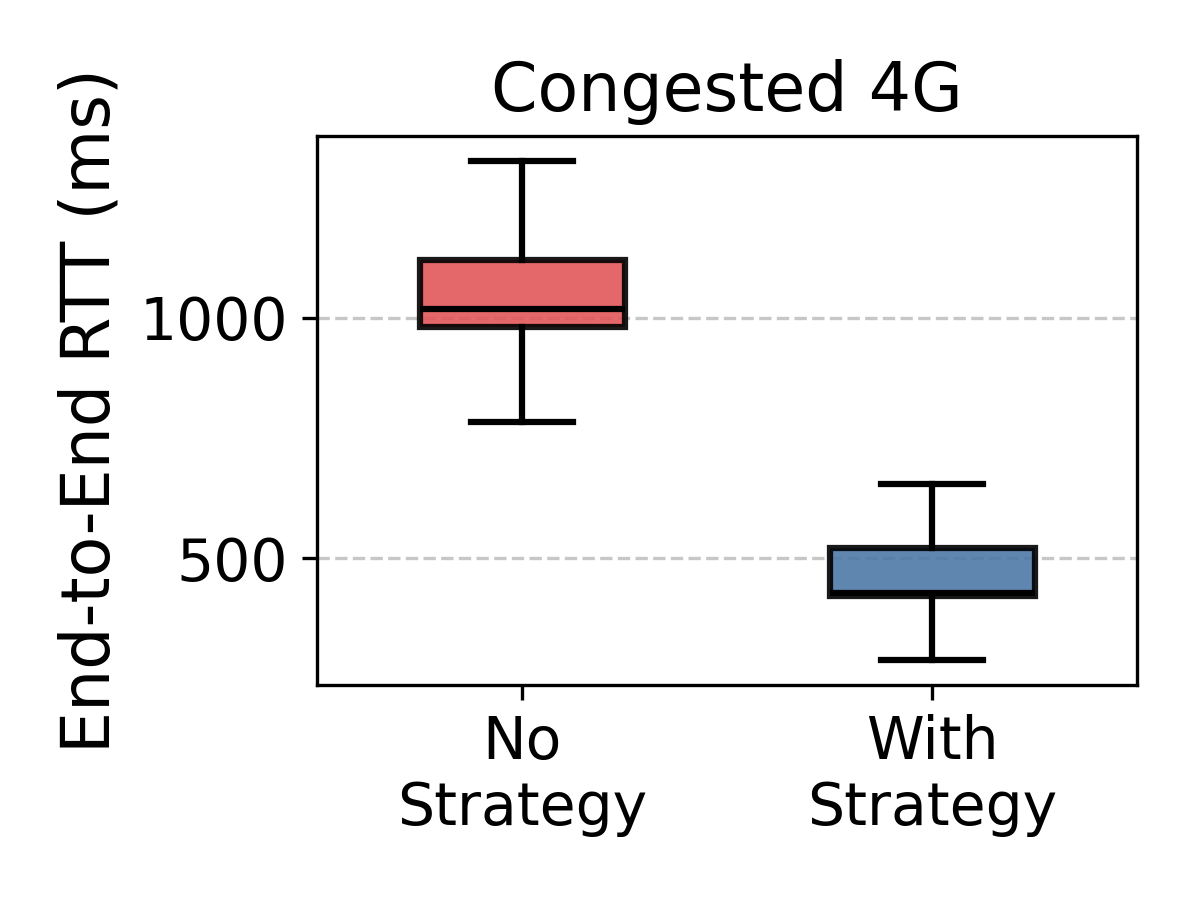}
    }
    \subfloat[4G\&5G Hybrid]{
        \includegraphics[width=0.19\textwidth]{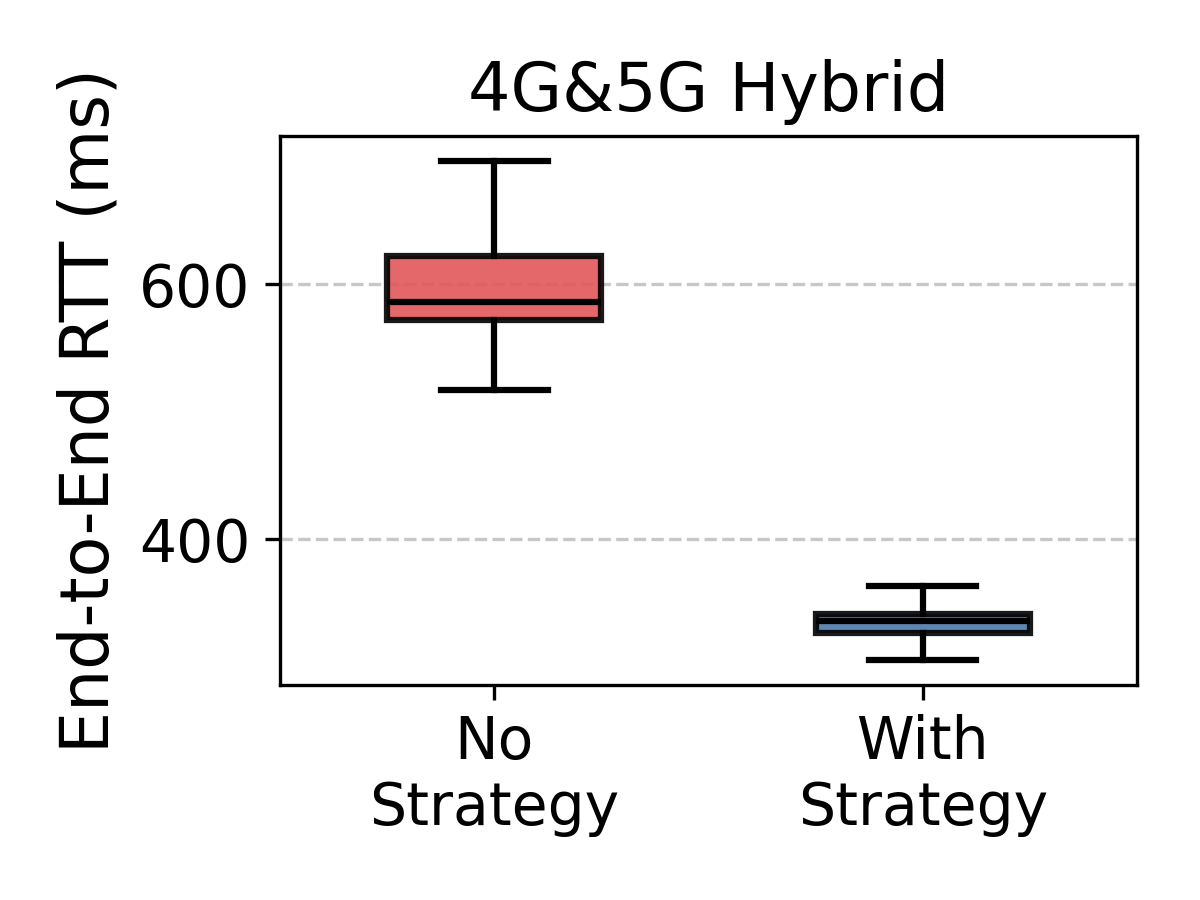}
    }
    \subfloat[Good 5G]{
        \includegraphics[width=0.19\textwidth]{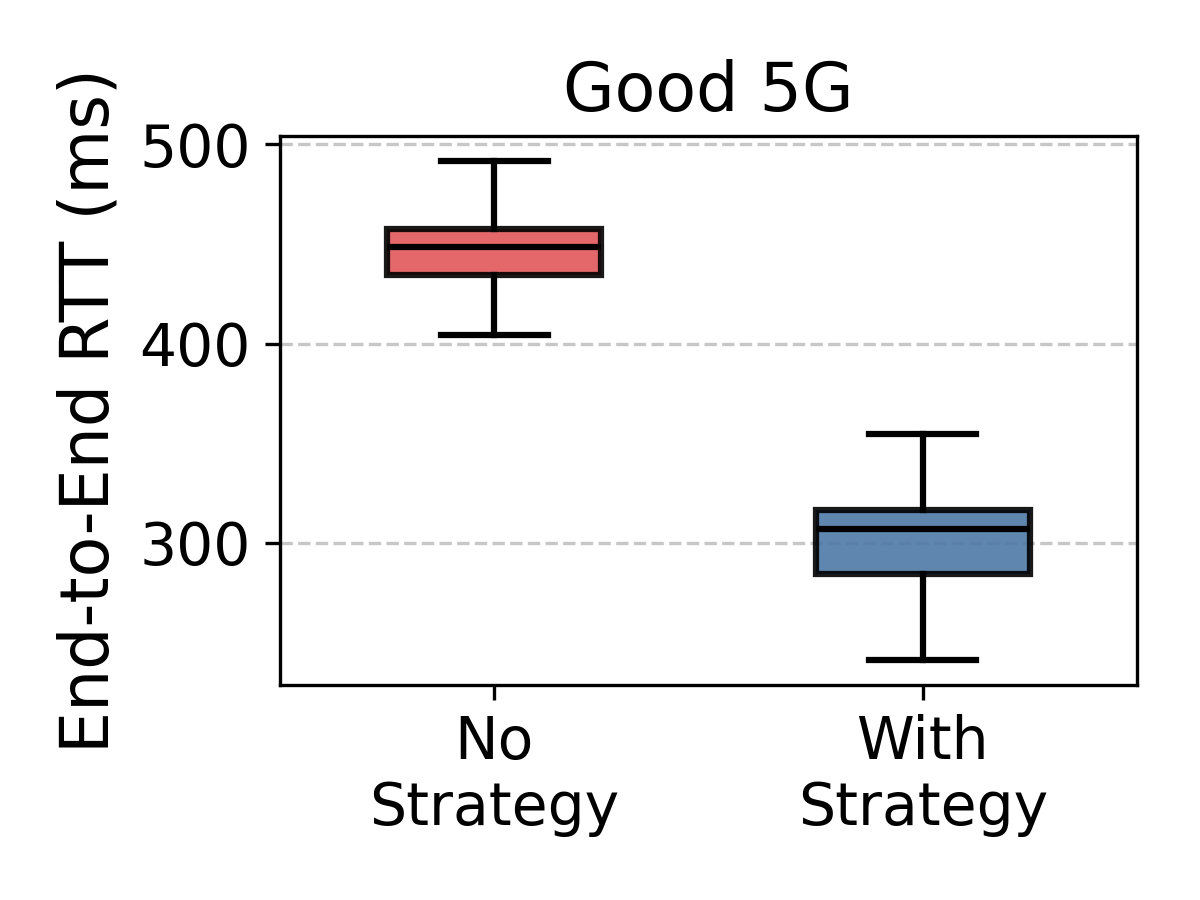}
    }
    \subfloat[Ultra-Smooth 5G]{
        \includegraphics[width=0.19\textwidth]{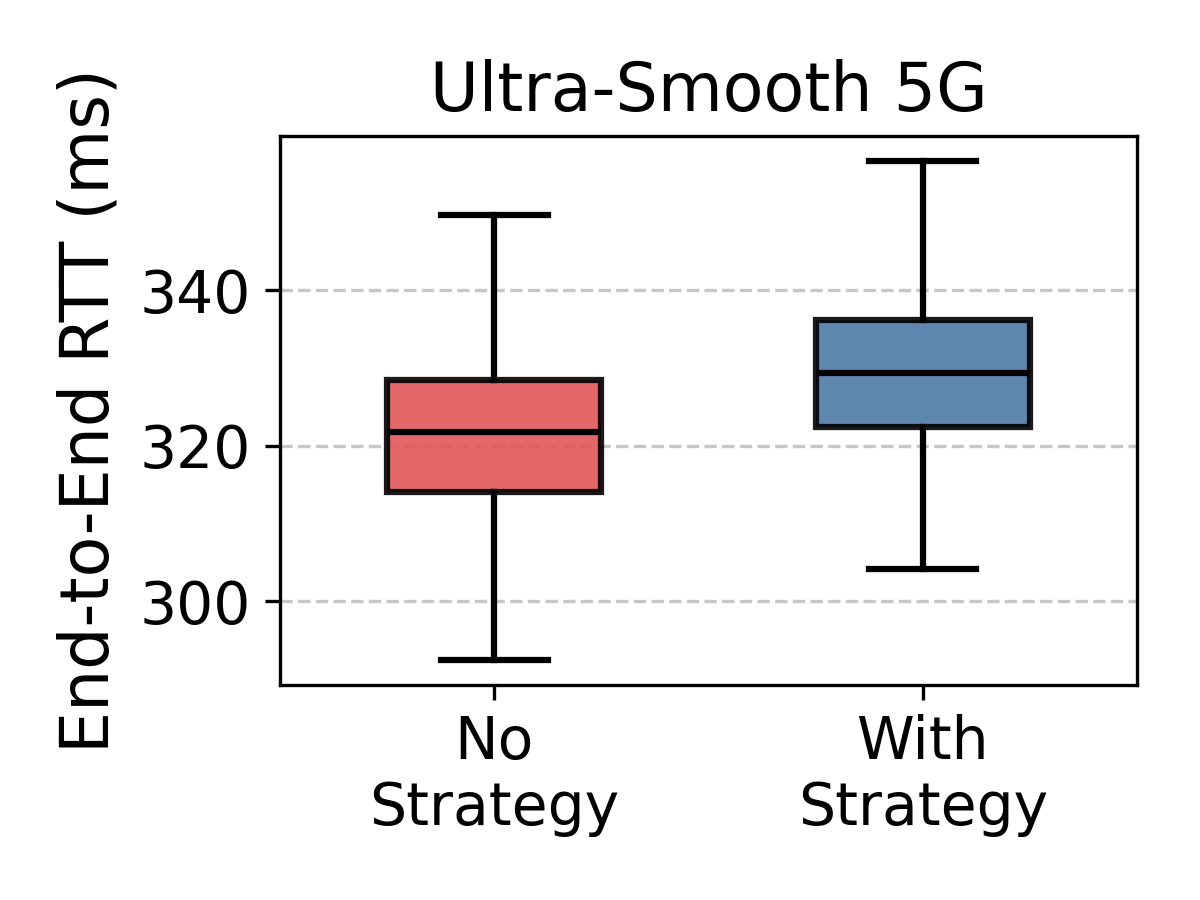}
    }
    \caption{End-to-end round-trip time (RTT) distributions under five simulated network conditions, comparing a static baseline with the proposed network-adaptive encoding policy.}
    \label{fig:rtt_comparison}
\end{figure*}

\subsection{Outcome Measures}

System performance was quantified along two complementary dimensions: 
(i) temporal responsiveness of the end-to-end processing loop, and 
(ii) fidelity of the returned semantic representation.
This separation reflects the central trade-off explored in this study between perceptual timing constraints and spatial detail.

\subsubsection{Temporal Responsiveness}

Temporal responsiveness was quantified using the end-to-end \ac{RTT} for each transmitted frame, measured from client-side transmission to receipt of the corresponding response. 
This metric captures the combined effects of network delay, queuing, encoding, decoding, and server-side inference.

In addition, mean server-side inference time was measured independently under each network scenario to characterize the computational contribution to end-to-end delay and to assess how adaptive downscaling influences remote processing time.

\subsubsection{Perceptual Fidelity Measures}

Because the present evaluation focuses on system-level behavior rather than human psychophysical performance, perceptual fidelity was assessed using established image-based measures applied to the semantic segmentation output. 
Specifically, we report the \ac{SSIM} and the \ac{BF} score.

\Ac{SSIM} quantifies preservation of global structural information in the returned representation relative to a reference, providing a proxy for scene-level perceptual integrity. 
The BF score captures boundary precision and is particularly relevant for prosthetic vision pipelines that rely on accurate delineation of salient objects and scene structure. 
Together, these measures enable systematic characterization of the latency-fidelity trade-off induced by network-adaptive encoding under varying connectivity conditions.

\section{RESULTS}

\subsection{Temporal Responsiveness Under Network Impairment}

We first evaluated the ability of the proposed network-adaptive encoding strategy to stabilize end-to-end latency under varying network conditions. 
System performance was assessed across five connectivity regimes ranging from severely constrained 4G-like conditions to near-ideal 5G-like operation, as defined in Table~\ref{tab:network_conditions}.

Figure~\ref{fig:rtt_comparison} shows the distribution of end-to-end \ac{RTT} measured for each network scenario, comparing the static baseline configuration with the proposed adaptive policy.

Under severely constrained network conditions, including Extreme Congested 4G and Congested 4G, the adaptive policy substantially reduced end-to-end \ac{RTT} relative to the static baseline. 
In these regimes, median \ac{RTT} was reduced by approximately 60--70\%, reflecting the controller’s ability to prevent queue buildup and excessive delay accumulation by dynamically reducing spatial resolution and transmission rate.

As network conditions improved, the difference between adaptive and static configurations diminished. 
In the Ultra-Smooth 5G condition, both configurations exhibited low and stable \ac{RTT}, and the adaptive policy introduced a small additional overhead due to monitoring and reconfiguration logic. 
This behavior is expected, as adaptation provides limited benefit when bandwidth and latency constraints are negligible.

\begin{figure}[!tb]
    \centering
    \includegraphics[width=0.48\textwidth]{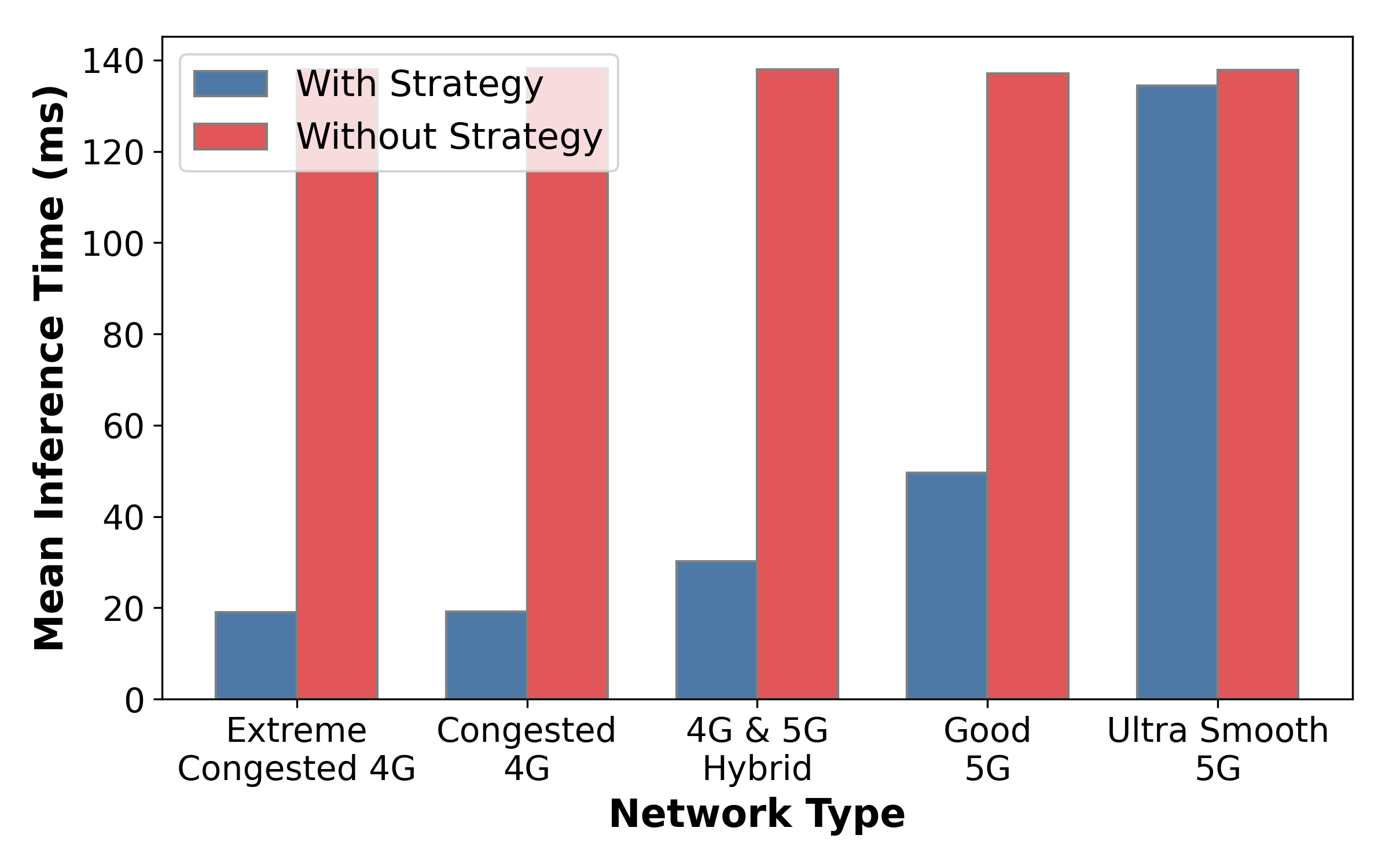}
    \caption{Mean server-side inference time under each network condition for static and adaptive configurations.}
    \label{inference_time}
\end{figure}

\subsection{Server-Side Inference Time}

To isolate the computational contribution to end-to-end delay, we additionally measured mean server-side inference time under each network scenario (Fig.~\ref{inference_time}).

During periods of network congestion, adaptive downscaling substantially reduced inference time by limiting the spatial resolution of frames processed by the cloud server. 
In the Extreme Congested 4G condition, mean inference time decreased from approximately 118\,ms under the static baseline to 19\,ms with adaptation. 
This reduction contributes directly to improved temporal responsiveness and helps maintain synchronization between visual updates and user motion.

Under high-quality network conditions, inference time differences between configurations were minimal, consistent with the controller operating in its highest-fidelity regime.
These inference-time reductions should be interpreted as input-dependent computational savings rather than as changes in model complexity. 
The number of PIDNet parameters and the network architecture were identical across static and adaptive conditions. 
However, the adaptive policy changes the effective computational load presented to the model by reducing the spatial dimensions and compression quality of the input frame. 
Because segmentation inference cost scales with the number of processed pixels, downscaling under congested conditions reduces server-side computation in addition to reducing network payload size. 
Thus, the observed end-to-end latency reflects coupled effects of communication delay, queuing, frame encoding, and input-dependent model inference time.

\subsection{Perceptual Fidelity Trade-Offs}

We next examined how adaptive encoding impacts the fidelity of the returned semantic representation. 
Perceptual fidelity was quantified using \ac{SSIM} and \ac{BF} scores, as summarized in Table~\ref{tab:model_accuracy}.

Across all network conditions, adaptive encoding resulted in modest reductions in \ac{SSIM} relative to the static baseline. 
Even under the most severe network impairment, \ac{SSIM} declined by only 3.14\%, indicating that global scene structure was largely preserved despite aggressive compression and downscaling.

As network conditions improved, \ac{SSIM} values under the adaptive policy converged toward the static baseline, reaching near-identical performance in the Ultra-Smooth 5G regime.

In contrast, \ac{BF} scores exhibited substantially greater sensitivity to adaptive downscaling. 
Under Extreme Congested 4G conditions, \ac{BF} score decreased from 50.38\% in the static baseline to 17.30\% with adaptation, reflecting the loss of fine-grained boundary detail at low spatial resolutions.

\Ac{BF} scores increased monotonically with improving network quality, reaching parity with the static baseline under Ultra-Smooth 5G conditions. 
This pattern highlights a key trade-off: during network congestion, the system prioritizes temporal continuity at the expense of precise boundary delineation.

Taken together, these results demonstrate that the proposed adaptive policy effectively regulates the latency–fidelity trade-off inherent to cloud-assisted visual preprocessing. 
Under constrained network conditions, the system transitions into a low-fidelity but temporally responsive operating regime, substantially reducing end-to-end delay while preserving global scene structure. 
As network conditions improve, the controller automatically restores higher spatial fidelity without manual intervention.

\begin{table}[!tb]
    \centering
    \caption{Perceptual fidelity measures under static and adaptive configurations across network conditions.}
    \renewcommand{\arraystretch}{1.2}
    \begin{tabular}{lcccc}
        \toprule
        \textbf{Scenario} 
        & \multicolumn{2}{c}{\textbf{SSIM (\%)}} 
        & \multicolumn{2}{c}{\textbf{BF Score (\%)}} \\
        \cmidrule(lr){2-3} \cmidrule(lr){4-5}
        & \textbf{Adaptive} & \textbf{Static} 
        & \textbf{Adaptive} & \textbf{Static} \\
        \midrule
        Extreme Congested 4G & 78.60 & 81.74 & 17.30 & 50.38 \\
        Congested 4G         & 78.67 & 81.74 & 17.99 & 50.38 \\
        4G\&5G Hybrid        & 79.84 & 81.74 & 27.81 & 50.38 \\
        Good 5G              & 80.76 & 81.74 & 39.01 & 50.38 \\
        Ultra-Smooth 5G      & 81.78 & 81.74 & 50.70 & 50.38 \\
        \bottomrule
    \end{tabular}
    \label{tab:model_accuracy}
\end{table}

\section{DISCUSSION}

This work examines the feasibility of cloud-assisted visual preprocessing for future visual neuroprostheses under realistic network constraints. 
By framing remote inference as a perceptually constrained control problem rather than a static computer vision pipeline, we demonstrate that adaptive visual encoding can substantially mitigate the impact of network-induced delays. 
Across a range of simulated wireless conditions, the proposed closed-loop strategy stabilizes end-to-end latency by dynamically trading spatial fidelity for temporal continuity, preserving a usable perceptual stream even under severe congestion.

From a neural engineering perspective, the key contribution of this study is not a specific vision model or network protocol, but a principled systems-level framework for reasoning about timing, fidelity, and network variability in artificial vision pipelines. 
Our results delineate operating regimes in which cloud-based preprocessing may remain viable for visual neuroprostheses, and identify perceptual trade-offs that must be managed to maintain safety and usability.

\subsection{Temporal Continuity as a Primary Design Constraint}

A central takeaway from this study is that temporal continuity should be treated as a central design constraint for cloud-assisted artificial vision. 
Under network congestion, static cloud pipelines exhibit rapidly increasing end-to-end delay, which can disrupt the temporal consistency of the delivered neural stimulus. 
In contrast, the proposed adaptive policy explicitly regulates latency by limiting queue buildup and inference time, enabling the system to maintain a minimum update rate even when bandwidth and packet loss are unfavorable.

Importantly, the observed reductions in latency were achieved with only modest degradation of global scene structure, as reflected by relatively stable \ac{SSIM} values across network regimes. 
This suggests that, for mobility-oriented tasks, preserving coarse scene layout and temporal alignment may be more critical than maintaining fine-grained spatial detail when network resources are limited.

\subsection{Latency-Fidelity Trade-Offs in Artificial Vision}

The results also highlight an inherent latency-fidelity trade-off in cloud-assisted preprocessing. 
While adaptive downscaling effectively reduces latency, it disproportionately impacts boundary precision, as reflected by reductions in \ac{BF} score under severe congestion. 
This finding is consistent with the role of high-frequency spatial information in semantic segmentation and underscores that different perceptual attributes degrade at different rates under adaptation.

From a systems design standpoint, this asymmetry suggests that adaptive policies should be task-aware. 
For example, navigation and obstacle avoidance may tolerate reduced boundary precision if temporal alignment is preserved, whereas tasks such as object identification or reading may require higher spatial fidelity and thus different adaptation strategies. 
These considerations point toward the need for context-dependent control policies that modulate adaptation based not only on network state but also on behavioral goals.

\subsection{Model Architecture and Computational Trade-Offs}

The present study used PIDNet as a fixed real-time semantic segmentation backbone to evaluate the feasibility of network-adaptive cloud preprocessing. 
This choice affects the absolute latency values reported here, because different segmentation architectures impose different computational loads, memory requirements, and sensitivity to input resolution. 
Large segmentation models may improve semantic accuracy or robustness in complex scenes, but their higher computational cost would increase server-side inference time and make the system more vulnerable to network-induced delay. 
Conversely, very lightweight mobile architectures may reduce inference time but could sacrifice boundary precision or semantic richness, both of which may be important for prosthetic vision~\cite{beyeler_towards_2022,han_deep_2021}.

For this reason, the results should be interpreted as demonstrating a systems-level control principle rather than identifying a single optimal segmentation model. 
The adaptive policy operates upstream of the model by changing the visual encoding parameters sent to the cloud. 
As a result, the same framework could be paired with alternative segmentation backbones, including lightweight edge models, transformer-based models, or task-specific neuroprosthetic preprocessing models. 
Future work should directly compare such architectures under matched network impairments to determine how model size, parameter count, input resolution, and segmentation fidelity jointly constrain usable cloud-assisted artificial vision.

\subsection{Limitations and Future Directions}

Several limitations of the present study warrant consideration. 
First, perceptual fidelity was assessed using algorithmic measures rather than human psychophysical evaluation. 
While SSIM and BF score provide useful proxies for global structure and boundary integrity, future work should incorporate behavioral experiments with prosthesis users or sighted participants viewing simulated artificial vision to more directly assess perceptual consequences.

Second, the current implementation modulates semantic detail indirectly through input degradation. 
Future systems could extend this framework by explicitly requesting coarse or fine semantic outputs from the cloud inference service, enabling more targeted control over perceptual content without relying solely on spatial downscaling. 
Additionally, hybrid architectures that combine lightweight on-device preprocessing with cloud-based semantic labeling may further improve robustness under extreme network conditions.

Finally, the adaptive policy explored here relies on discrete operating regimes and a single network feedback signal. 
More sophisticated controllers that incorporate predictive models of network variability or additional feedback signals could enable smoother transitions and improved performance.

Despite these limitations, the present results provide a concrete foundation for exploring cloud-assisted artificial vision as a viable design space rather than an all-or-nothing proposition.

\section{CONCLUSION}

As visual neuroprostheses increasingly incorporate advanced machine learning–based preprocessing, computational demands are likely to outpace what can be supported on battery-powered devices alone. 
This study shows that cloud-assisted preprocessing, when coupled with network-aware adaptive encoding, can remain compatible with perceptual timing constraints critical for safe and effective artificial vision. 
By grounding cloud inference in a control-theoretic and perceptually informed framework, this work contributes a systems-level perspective that may inform the design of future neuroprosthetic platforms bridging computation, communication, and neural stimulation.

\bibliographystyle{ieeetr}
\bibliography{references}

\end{document}